\documentclass[journal]{IEEEtran}  % Comment this line out
\IEEEoverridecommandlockouts                              % This command is only
\usepackage[utf8]{inputenc}
\usepackage[T1]{fontenc}
\usepackage{graphicx}
\usepackage{float}
\usepackage{url}
\usepackage{siunitx}
\usepackage{subcaption}
\usepackage{xcolor}
\usepackage{comment}
\usepackage{hhline}
\usepackage{soul}
\usepackage{flushend}
\usepackage[font={small}]{caption}
\title{An RRAM-Based Implementation of a Template Matching Circuit for Low-Power Analogue Classification}

\begin{document}

\author{Patrick Foster, Georgios Papandroulidakis, Alex Serb,~\IEEEmembership{~Senior Member,~IEEE}, \\Spyros Stathopoulos,~\IEEEmembership{~Member,~IEEE}, and Themis Prodromakis,~\IEEEmembership{~Senior Member,~IEEE}\\%
Centre for Electronics Frontiers (CEF), Institute for Integrated Micro and Nano Systems, \\University of Edinburgh, UK
\thanks{P. Foster, G.Papandroulidakis, A. Serb S. Stathopoulos and T. Prodromakis are with the Centre for Electronics Frontiers, University of Edinburgh, UK\\
E-mails: \{pfoster4, gpapandr, aserb, s.stathopoulos, t.prodromakis\}@ed.ac.uk\\
Contact e-mail: pfoster4@ed.ac.uk}% <-this % stops a space
\thanks{This work was supported in part by the Engineering and Physical Sciences Research Council (EPSRC) Programme under Functional Oxide Reconfigurable Technologies (FORTE) Grant EP/R024642/1 and in part by the RAEng Chair in Emerging Technologies under Grant CiET1819/2/93.}% <-this % stops a space
\thanks{}}

% The paper headers
%\markboth{IEEE Transaction on Circuits and Systems I: Regular Papers,~Vol., No., September 2024}%
%{P. Foster \MakeLowercase{\textit{et al.}}: A RRAM-Based Implementation of a Template Matching Circuit for Low-Power Analogue Classification}

\maketitle
%\thispagestyle{empty}
%\pagestyle{empty}

%%%%%%%%%%%%%%%%%%%%%%%%%%%%%%%%%%%%%%%%%%%%%%%%%%%%%%%%%%%%%%%%%%%%%%%%%%%%%%%%
\begin{abstract}

Recent advances in machine learning and neuro-inspired systems enabled the increased interest in efficient pattern recognition at the edge. A wide variety of applications, such as near-sensor classification, require fast and low-power approaches for pattern matching through the use of associative memories and their more well-known implementation, Content Addressable Memories (CAMs). Towards addressing the need for low-power classification, this work showcases an RRAM-based analogue CAM (ACAM) intended for template matching applications, providing a low-power reconfigurable classification engine for the extreme edge. The circuit uses a low component count at 6T2R2M, comparable with the most compact existing cells of this type. In this work, we demonstrate a hardware prototype, built with commercial off-the-shelf (COTS) components for the MOSFET-based circuits, that implements rows of 6T2R2M employing $TiO_{x}$-based RRAM devices developed in-house, showcasing competitive matching window configurability and definition. Furthermore, through simulations, we validate the performance of the proposed circuit by using a commercially available 180nm technology and in-house RRAM data-driven model to assess the energy dissipation, exhibiting $\SI{60}{\pico\joule}$ per classification event.

\end{abstract}


%%%%%%%%%%%%%%%%%%%%%%%%%%%%%%%%%%%%%%%%%%%%%%%%%%%%%%%%%%%%%%%%%%%%%%%%%%%%%%%%
\section{Introduction}\label{sec:intro}

\IEEEPARstart{I}{N} recent decades, the conventional von Neumann architecture that is dominating most modern computing systems seems to be reaching a performance saturation point mainly due to the well-documented memory wall bottleneck occurring as a result of the separation between computing units and memory modules \cite{Wilkes1995}\cite{McKee2004}. As further miniaturisation of MOSFET technology provides diminishing performance gains with every new CMOS generation, a lot of focus has been placed on investigating novel computing technologies and architectures to satisfy the increasing need for higher performance and low energy computing. This is especially important for power and area restricted applications at the edge and extreme edge \cite{Menon2021}. One important avenue of investigation is the neuro-inspired architectures, where the close integration of computing and memory is observed \cite{Graves2020}. This co-location of computing and memory enables the implementation of novel in-memory (IMC) or near-memory (NMC) computing approaches that showcase promising results in continuing the performance improvements in critical algorithms, such as machine learning (ML), while being also energy efficient \cite{Sebastian2020}. One such paradigm is the introduction of associative memories that can perform template matching operation, a feature critical for signal processing at the edge where noisy inputs require fast and low power classification. \par
%===================================================================
One of the most well-documented models of neuro-inspired associative memories are the content addressable memories (CAMs). CAM is a broad term describing memory-centric technologies where the memory cells compare their stored state with an input locally and in parallel. This allows a bank of CAM to be quickly searched by applying the information that is being sought to the input of the array and monitoring the cell outputs \cite{TXLpaper}. The combining of low level computing tasks with memory blurs the boundary between the two and has the potential for improvements in computing \cite{CAMtrends} especially in ML applications for resource-constrained implementations. CAM can serve as the foundation for novel Artificial Neural Nets (ANNs) and other neuromorphic systems especially when co-integrated with emerging memory technologies, such as Resistive RAM (RRAM) \cite{treeBasedMLCAM}.\par
%===================================================================
    \begin{figure}[t]
        \centering
        \includegraphics[width=0.8\linewidth]{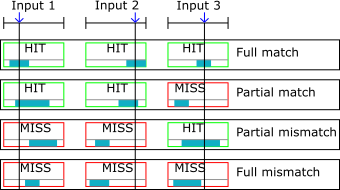}
        \caption{Concept schematic showcasing an example analogue content addressable memory (ACAM) parallel search operation. Each cell performs a comparison between an input of the analogue template query and its match window, as defined by its memory contents. If all cells in a row have a hit then the ACAM responds with the memory address of the full matching row. The match/mismatch can be define according to an array-wide threshold.}
        \label{fig:CAM diagram}
    \end{figure}
%===================================================================
In this work, we are showcasing a RRAM-CMOS ACAM circuit capable of mapping two analogue thresholds to identify a specific matching window. We provide detailed simulations and proof-of-concept experimental results by integrating in-house developed $TiO_{x}$-based RRAM devices. The proposed ACAM cell acts as reconfigurable building block for a template matching systems, which performs low-power operation for associative inference at the edge. The reconfigurability is introduced to the circuit through the RRAM technology. We provide a detailed analysis through SPICE simulations of the proposed circuit to assess its performance and then develop a proof-of-concept experimental board using both Commercial off the shelf (COTS) and experimental RRAM devices to further investigate the behaviour of such template matching system. More specifically, in Section \ref{sec:background}, we provide a brief overview of existing RRAM ACAM technologies and their applications. In Section \ref{sec:circ}, we explain the circuit design and operation. In Section \ref{sec:sims}, we are showing simulation results of the novel split TXL circuit design using a commercially available 180nm technology. In Section \ref{sec:mes}, we are exhibiting hardware results, and in Section \ref{sec:disc}, we are providing concluding remarks. \par
%===================================================================

\section{Background}\label{sec:background}
In general, CAMs can be divided into two types: digital CAM and analogue CAM (ACAM). Digital CAM stores only binary information and compare this stored value with a binary query vector \cite{binCAM}. In some cases, CAM can have a third stored state for 'don't care'; this is called ternary CAM \cite{Graves2020}\cite{ternCAM}. SRAM-based CAM systems, either binary or ternary, are well-known systems and are commonly found deployed as network switches, where they store large tables of MAC addresses for fast signal routing \cite{camTable}. While CMOS binary CAM is well established, there is ongoing research in the area, trying to integrate emerging non-volatile memory (NVM) elements, such as RRAM \cite{contAddrCell}. Given the multi-bit memory capabilities of these emerging memory technologies, the rapid development of novel ACAM design has been exhibited. \par
%===================================================================
ACAM design follows the main principles of digital CAM design but with appropriate circuit modifications to enable the in-situ processing of analogue signals. This is usually achieved by using multi-bit memory cells. In addition to their use in neuromorphic systems \cite{neuroCAM}, this type of CAM can also be used in analogue signal analysis methods such as template matching \cite{templateCAM}. In this application, ACAM cells are organised as memory arrays with each memory cell having a small additional circuitry for a per cell comparison operation between an input and the stored value \cite{TXLpaper}\cite{treeBasedMLCAM}. The ACAM array performs a massively parallel search operation for all available templates inside the memory and responds with the address of the classified template if a match event occurs. Unlike binary CAM research, where the goal is to add nonvolatility to the cells without increasing power dissipation, analogue CAM research seeks to replace high power dissipation and area comparators with simplified circuits that take advantage of the analogue nature of emerging memory technologies. The advent of emerging memory technologies enables the implementation of CAMs with analogue memory components making possible the usage of such systems to perform template matching in the analogue domain without the need to convert information from analogue to digital \cite{TXLpaper}. While there are attempts to use other emerging memory technologies such as FeRAM \cite{ferroCAM}\cite{ferroCAM2} for this purpose, RRAM implementations of ACAM can showcase advantages due to the granularity of conductance states in RRAM technologies and the ease of tuning offered by resistive elements \cite{rramCAM}\cite{rramCAM2}\cite{Pedretti2021}. Such implementations offer a promising candidate for near-sensor fast template matching engines for a new era of bio-signal processing systems at the extreme edge. \par
%===================================================================
Recently, a range of different ACAM cell designs employing emerging memory technologies have been proposed towards implementing novel template matching systems. In \cite{TXLpaper}, a 6T2R design has been shown to implement a template matching circuit capable of detecting specific voltage levels using a 2T2R and a 4T memory readout circuit. This design showcase good area utilisation and showcase an early look on how hybrid RRAM-CMOS circuits can be used to map a threshold point for which a short event-triggered pulse is produced. In \cite{hpCAM}, a $Ta/TaO_{x}$-based ACAM design using also a 6T2R configuration has been showcased. In this design, two different thresholds are mapped in the 2T2R circuit which is implemented as two separate 1T1Rs. The rest of the cell (4T) implements the window matching comparison. The circuit shows good area and match window configurability but the use of 1T1R voltage dividers when reading the memory could result in increased power consumption. \par
%===================================================================
%===================================================================
\section{Circuit Design and Operation}\label{sec:circ}
\begin{figure}[ht]
        \centering
        \includegraphics[width=0.9\linewidth]{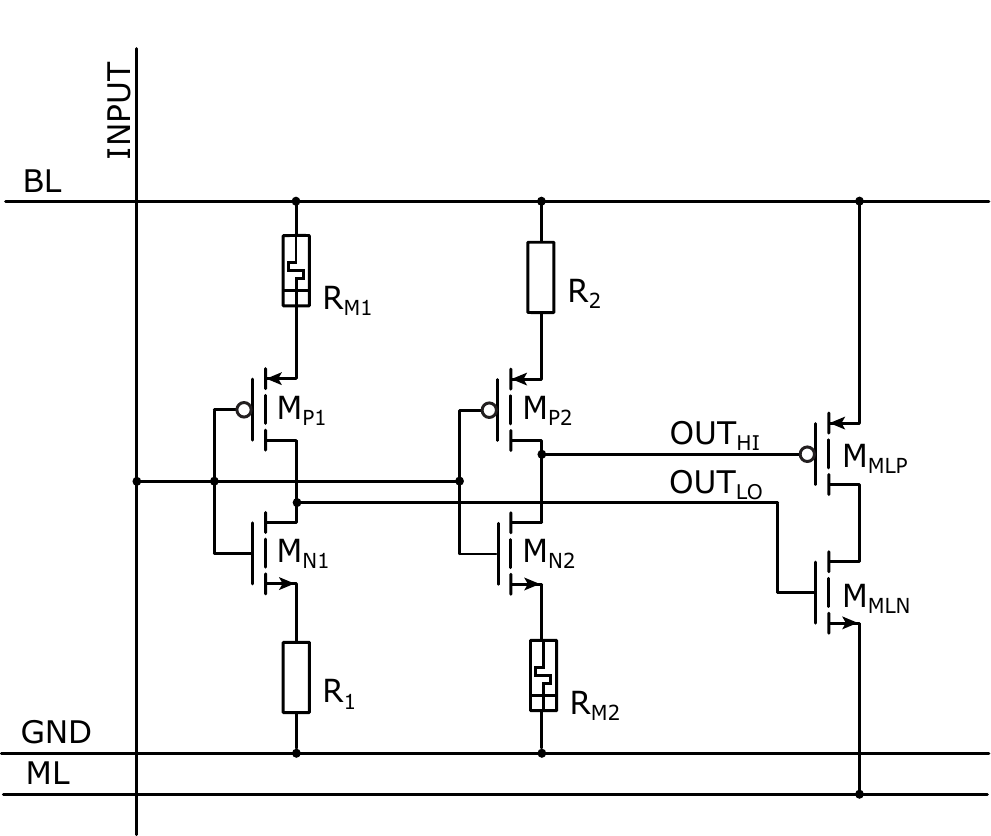}
        \caption{Circuit schematic of the split TXL ACAM cell proposed in this work. The circuit implements two hybrid RRAM-CMOS inverters ($R_{M1}$-$M_{P1}$-$M_{N1}$-$R_{1}$ and $R_{2}$-$M_{P2}$-$M_{N2}$-$R_{M2}$) that map the low and high thresholds (encoded as the outputs $OUT_{LO}$ and $OUT_{HI}$ of the hybrid inverters) and two extra MOSFET devices ($M_{PML}$ and $M_{NML}$) that performs the comparison between the low and high threshold and connect the output with the $BL$ charging the matchline $ML$ for the case of match. The $BL$ has been precharged to $V_{DD}$ before the ACAM operation. Devices $R_{1}$ and $R_{2}$, that operate as the other part of the resistive ratio determining the matching window, are implemented as static resistive elements.}
        \label{fig:TXLdiagramAgain}
    \end{figure}

The circuit proposed in this work, shown in Fig. \ref{fig:TXLdiagramAgain}, is based upon a skewed inverter design showcased in \cite{TXLpaper}. The cell design in \cite{TXLpaper} uses a hybrid RRAM-CMOS inverter to adjust the switching point of a second inverter stage, with a current mirror to detect the higher current at that switching point. By comparison, our proposed design uses two hybrid inverters to map two analogue thresholds to enable a higher configurability of the matching window. Two complementary FET pairs with source degeneration are used as the input stage with their outputs assessed by an XOR mode transmission gate. The current through this gate feeds provides the output signal, which can be connected in parallel with other cells to a simplified matchline pull-up network to perform template matching. The ratio between source degeneration on the input PMOS and NMOS alters the threshold of the inverter. As the threshold is dependent on the ratio between the top and bottom degeneration resistors and not the total resistance of the branch, the threshold can be controlled by altering the state of only one of resistive elements. As such, this design uses a single dynamic RRAM device for each hybrid inverter to control each threshold, with the opposing degeneration being provided by a static balancing resistor. \par
%===================================================================
%===================================================================
When the input is above the threshold of the $R_{M1}$ branch inverter and below the threshold of the  $R_{M2}$ branch inverter, both the PMOS and NMOS of the output stage ($M_{MLP}$ and $M_{MLN}$, respectively) conduct, thus connecting the charged $BL$ to discharged $ML$. The $BL$ is charged to $V_{DD}$ and the $ML$ is discharged to $GND$. This disrupts the conventional operation of precharged $ML$ that remains charged in the case of a full row match or drains for the case of a single mismatch. Instead, this method enables the charge of the discharged $ML$ by a packet of charge controlled by the output branch of each cell in a row. This technique provides greater flexibility in the analogue operation of the row since we can define how many charge packets are required for a match and decide on a configurable threshold. This threshold is implemented as part of the matchline readout circuitry (i.e., the threshold voltage of a sense amplifier). While this configuration permits the independent control of each threshold with a single programmable resistive component, the required range of memristor states is quite high, although not to the degree required by previous works \cite{hpCAM} where the ratio of minimum to maximum resistive state was $5\times$ higher. This is because the configurable element must go both above and below the balancing resistor by a factor of 10 to achieve a fully configurable range of threshold values that could be helpful in supporting a wide range of applications. \par

\section{Hardware Measurement}\label{sec:mes}

\begin{figure}[t]
        \centering
        \begin{subfigure}[t]{\linewidth}
            \centering
            \includegraphics[trim=0cm 0cm 25cm 0cm, clip=true, angle=90, width=0.65\linewidth]{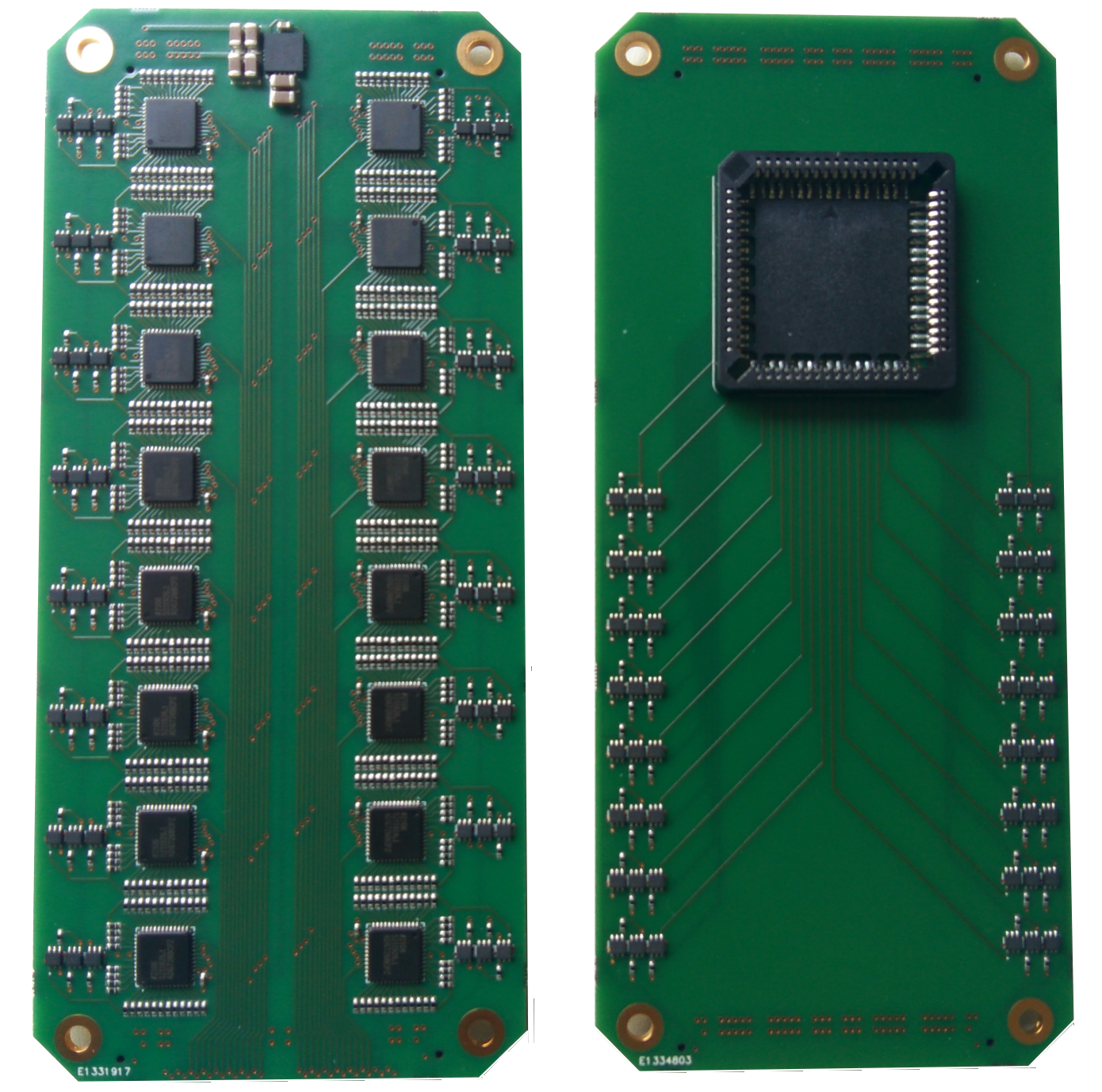}
            \caption{}
            \label{fig:TXLresDB}
        \end{subfigure}
        \begin{subfigure}[t]{\linewidth}
            \centering
            \includegraphics[trim=25cm 0cm 0cm 0cm, clip=true, angle=90, width=0.65\linewidth]{Pictures/TXL_dbs.png}
            \caption{}
            \label{fig:TXLmemDB}
        \end{subfigure}
        \caption{Photographs of the PCB prototypes. a) The initial PCB prototype using conventional resistor components for implementing rows of the proposed ACAM cell, b) The second RRAM-MOSFET PCB prototype implementing rows of the proposed ACAM cell with the connectivity to the resistor components being replaced by a PLCC68 socket for connecting experimental RRAM packages.}
        \label{fig:txlPCBsPhoto}
    \end{figure}

%Although there is a number of RRAM models, most existing models are limited in scope and detail and capture only some of the intrinsic behaviour of said devices. Analysis of this circuit must therefore be done with a physical model. In theory, the state of the memristors used should only matter at two points: the current at which the inverter switches, and at zero current. Because of this, models of CAM cell using linear resistors should be a reasonable approximation of the intended circuit, but this assumption must be tested. To this end, 
To experimentally test the TXL proof-of-concept we are exploring the design space of the circuits initially by using ideal resistors and then testing with in-house developed RRAM devices. Two CAM array PCBs were designed for use with an existing instrument platform \cite{arctwo}. The PCB implementations of this cell use the SSM6L36TU,LF \cite{SSM6L36TULF}, along with $\SI{1}{\mega\ohm}$ balancing and output resistors. The FETs used are small signal MOSFETs with threshold voltages comparable to integrated FETs. The high value balancing resistor results in low supply current in the RRAM-skewed inverters, allowing the inverters to switch a little closer to the supplies, reducing the required headroom. For the dynamic element, the resistor model (shown in Fig. \ref{fig:TXLresDB}) used a multiplexer switched array of 16 resistors, geometrically spaced between $\SI{100}{\kilo\ohm}$ and $\SI{10}{\mega\ohm}$. The RRAM PCB design (shown in Fig. \ref{fig:TXLmemDB}) used a PLCC68 package of $30\times\SI{30}{\micro\meter}$ $TiO_{x}$-based RRAM devices \cite{monolayerTiO}.\par

%===================================================================
    \begin{figure}[ht]
        \centering
        \includegraphics[width=0.7\linewidth]{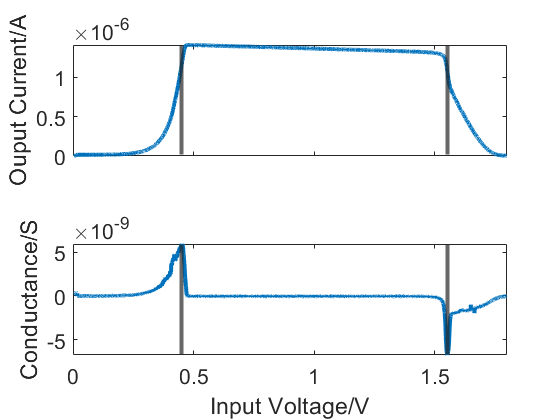}
        \caption{A diagram of the method used to determine window width. In this case the window width is calculated as $\SI{1.1}{\volt}$.}
        \label{fig:diffWidthDiagram}
    \end{figure}
%===================================================================
To assess how the window responded to changes in the control devices, the current output is denoised with a 50 sample moving average filter and the differential calculated. The window width is calculated as the difference in input voltage between the minimum and maximum of the derivative (Fig. \ref{fig:diffWidthDiagram}). This method was chosen as it provides a meaningful result in circumstances where the window peak is not consistent, where a more common method such as full width at half maximum (FWHM) might not. The differential method typically gives a slightly more pessimistic valuation of the circuit performance than FWHM.\par
%===================================================================
\subsection{Resistor-based Test PCB}\label{sec:resModel}

%===================================================================

The behaviour of the resistor-based emulator board was measured by setting one of the configurable elements to its maximum value and conducting an input sweep at each step of the other configurable element. In this PCB design, the CAM circuit was supplied at $\SI{1.8}{\volt}$ and the input swept from $\SI{0}{\volt}$ to the supply voltage by stepping through every DAC code. The current was measured at every step, for a total of 5898 samples. More information on how the DAC operates in our experimental setup using ArC TWO instrumentation board can be found in \cite{arctwo}.\par
%===================================================================
    \begin{figure}[ht]
        \centering
        \begin{subfigure}[b]{0.49\linewidth}
            \centering
            \includegraphics[width=\linewidth]{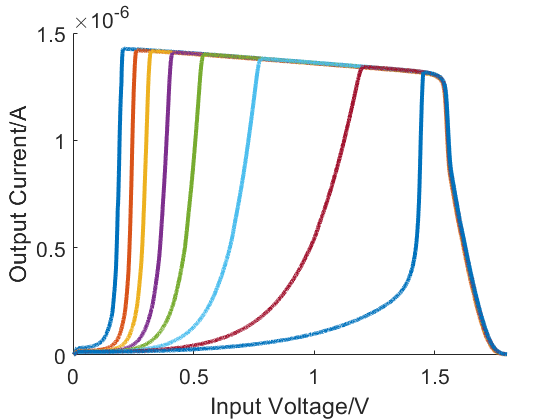}
            \caption{}
            \label{fig:resM1bThres}
        \end{subfigure}
        \begin{subfigure}[b]{0.49\linewidth}
            \centering
            \includegraphics[width=\linewidth]{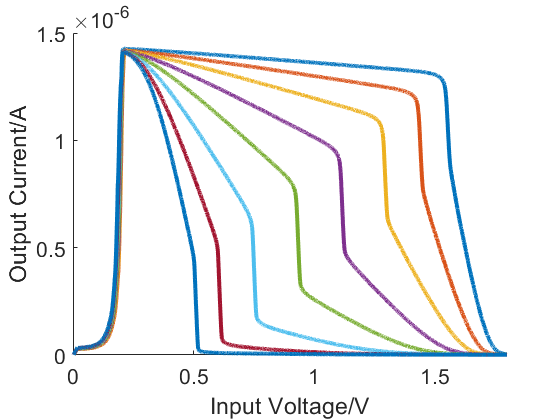}
            \caption{}
            \label{fig:M2tThres}
        \end{subfigure}
        \begin{subfigure}[b]{0.49\linewidth}
            \centering
            \includegraphics[width=\linewidth]{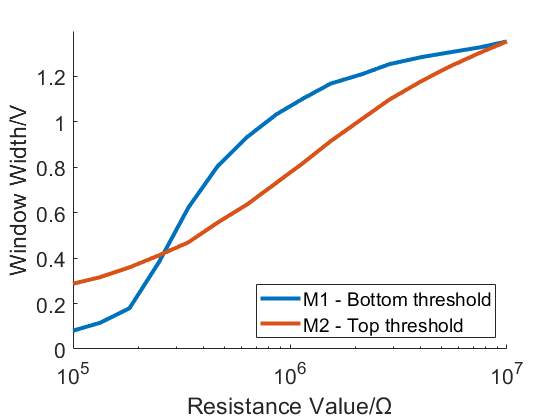}
            \caption{}
            \label{fig:resWidthState}
        \end{subfigure}
        \begin{subfigure}[b]{0.49\linewidth}
            \centering
            \includegraphics[width=\linewidth]{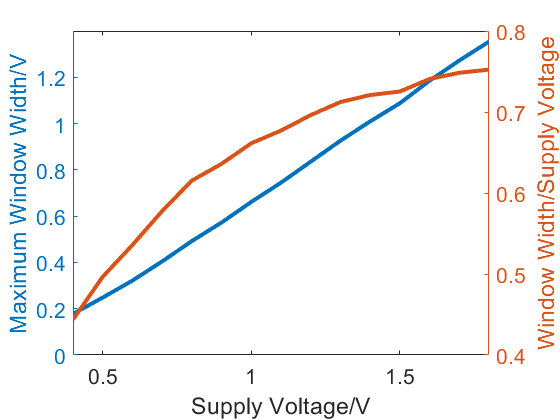}
            \caption{}
            \label{fig:resWidthPwr}
        \end{subfigure}
        \caption{Graphs showing the effect of M1/2 on the match window of the PCB resistor model. a) An IV sweep of M1 between $\SI{10}{\mega\ohm}$ and $\SI{100}{\kilo\ohm}$, showing that the threshold voltage increases as M1 drops. b) An identical sweep of M2, showing that the threshold voltage decreases as M2 drops. c) The width of the window as a function of resistor state. d) The maximum window width as a function of the supply voltage.}
        \label{fig:resModel}
    \end{figure}
%===================================================================
As can be seen in Fig. \ref{fig:resModel}, the circuit produces a significant output current at input values between the two thresholds. The upper threshold was measured to be almost linear with the logarithm of the control resistor. The upper threshold displays the characteristic behaviour of a current starved inverter, because the NMOS in the output branch acts as a voltage follower. This somewhat compromises the specificity of the circuit, but not to the extent that would justify the extra inverter needed to swap it for a PMOS. The lower threshold shows non-linearity behaviour of control, with a pronounced s-curve. Additionally, the maximum window width was calculated for a range of supply voltages. This was found to be linear, signifying that the headroom required is a function of the threshold value of the MOSFETs used, regardless of the operating current.\par
%===================================================================
\subsection{RRAM-based Test PCB}\label{sec:memModel}
\begin{figure}[ht]
        \centering
        \begin{subfigure}[b]{0.49\linewidth}
            \centering
            \includegraphics[width=\linewidth]{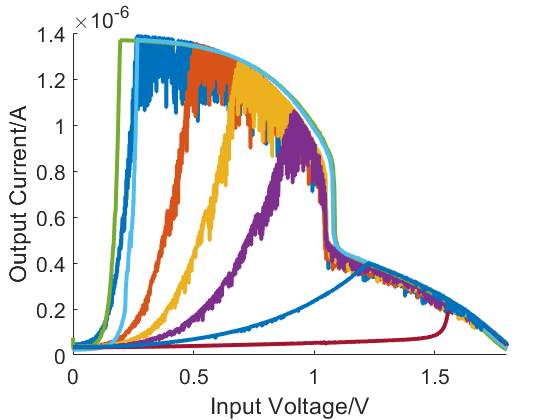}
            \caption{}
            \label{fig:resM1bThresMem}
        \end{subfigure}
        \begin{subfigure}[b]{0.49\linewidth}
            \centering
            \includegraphics[width=\linewidth]{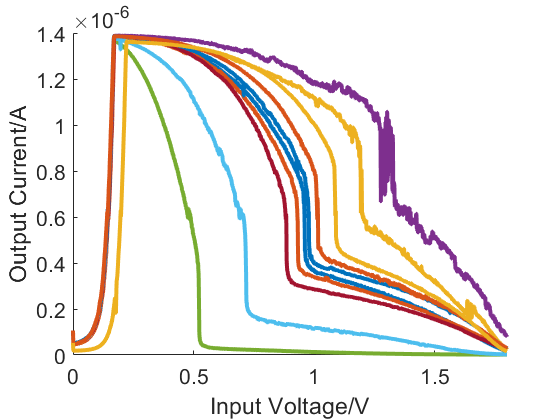}
            \caption{}
            \label{fig:M2tThresMem}
        \end{subfigure}
        \begin{subfigure}[b]{0.49\linewidth}
            \centering
            \includegraphics[width=\linewidth]{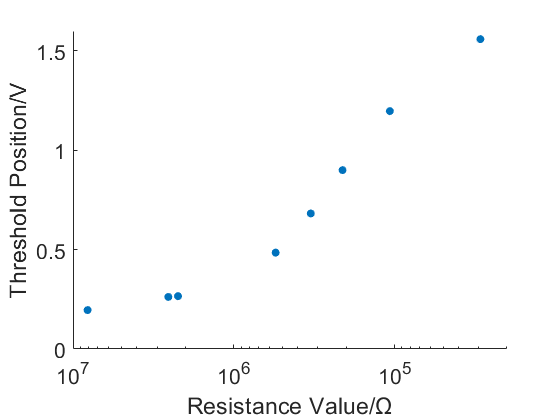}
            \caption{}
            \label{fig:bThresMem}
        \end{subfigure}
        \begin{subfigure}[b]{0.49\linewidth}
            \centering
            \includegraphics[width=\linewidth]{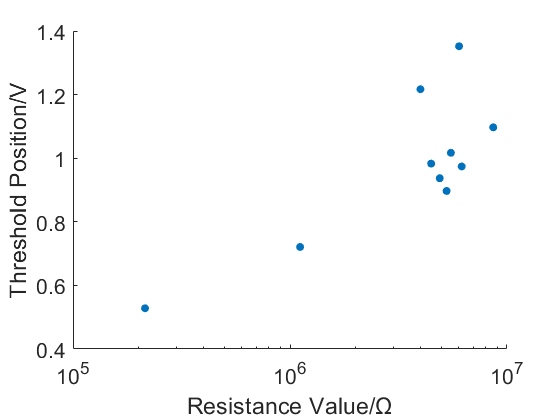}
            \caption{}
            \label{fig:tThresMem}
        \end{subfigure}
        \caption{Graphs showing the effect of M1/2 on the match window of the PCB RRAM model. a) An IV sweep of M1 between $\SI{8}{\mega\ohm}$ to $\SI{30}{\kilo\ohm}$ with M2 at approximately $\SI{7}{\mega\ohm}$, showing that the threshold voltage increases as M1 drops. b) A similar sweep of M2 from, $\SI{200}{\kilo\ohm}$ to $\SI{8.6}{\mega\ohm}$ with M1 at approximately $\SI{6}{\mega\ohm}$, showing that the threshold voltage decreases as M2 drops. c) The lower threshold voltage as a function of the M1 state. Note that the X axis is reversed compared to subfigure d). d) The upper threshold as a function of M2 state.}
        \label{fig:memModel}
    \end{figure}

%===================================================================
For the RRAM-based PCB tests, the RRAM devices were set to the preferred conductance state with $100-\SI{500}{\micro\second}$ pulses and then read at $\SI{250}{\milli\volt}$, before performing a sweep of the input voltage in $\SI{1}{\milli\volt}$ steps, reading the current at each step for a total of 1801 samples per trace. For the purposes of calculating the threshold position, the current output curve was denoised with the same 50 sample moving average filter. The mono-layer $TiO_{x}$-based RRAM devices used in this test \cite{monolayerTiO} are an experimental technology and exhibit some volatility and drifting of their conductance states. In the sweep of M1, some tests saw the memristor taking random values within a small range, as can be seen in Fig. \ref{fig:resM1bThresMem}. In the sweep of M2, a test at high resistive state saw the control device switching between two distinct values, as can be seen in the outermost trace of Fig. \ref{fig:M2tThresMem}. This may be indicative of the formation (and disruption) of conductive filaments within the titanium oxide layer of the RRAM devices, as this is one of the hypothesised mechanisms of the memristive behaviour of these devices \cite{memFilament}.\par
%===================================================================
%===================================================================
Toward avoiding issues with device-to-device and cycle-to-cycle stochasticity of RRAM devices, instead of assessing the window width, the position of each threshold was calculated. While this is less useful in characterising the circuit, it prevents issues with setting one threshold from affecting the accurate assessment of the other. Despite this, the tests on the bottom threshold showed a similar s-curve shape to the resistor model. The shape of the window produced is also very similar to the shape obtained in the resistor model. The upper threshold was not so well behaved, showing significant loss in the sharpness of the transition. Further, there was a loss of monotonicity at high resistive states, although this may be due to changes in the state of the M2 device between the measurement taken at the start of each test and the state it held during the test. Despite the irregularities, the traces produced in the tests of the memristor model are clearly of comparable shape and range of position to the results of the resistor model. Thus, we demonstrate that resistor tuned models of TXL circuits can be used as a good emulator for RRAM for the assessment of characteristics not explicitly related to the threshold shape, such as power consumption and footprint area. Through this method we are able to perform hardware experiments using conventional technologies and use the RRAM arrays in a program-once-read-many operational scheme, where the weights are defined through software and resistor-based emulator array testing and then mapped into a RRAM array.\par

 \subsection{RRAM-based Demonstrator Board}
\begin{figure*}[ht]
        \centering
        \includegraphics[width=0.92\linewidth]{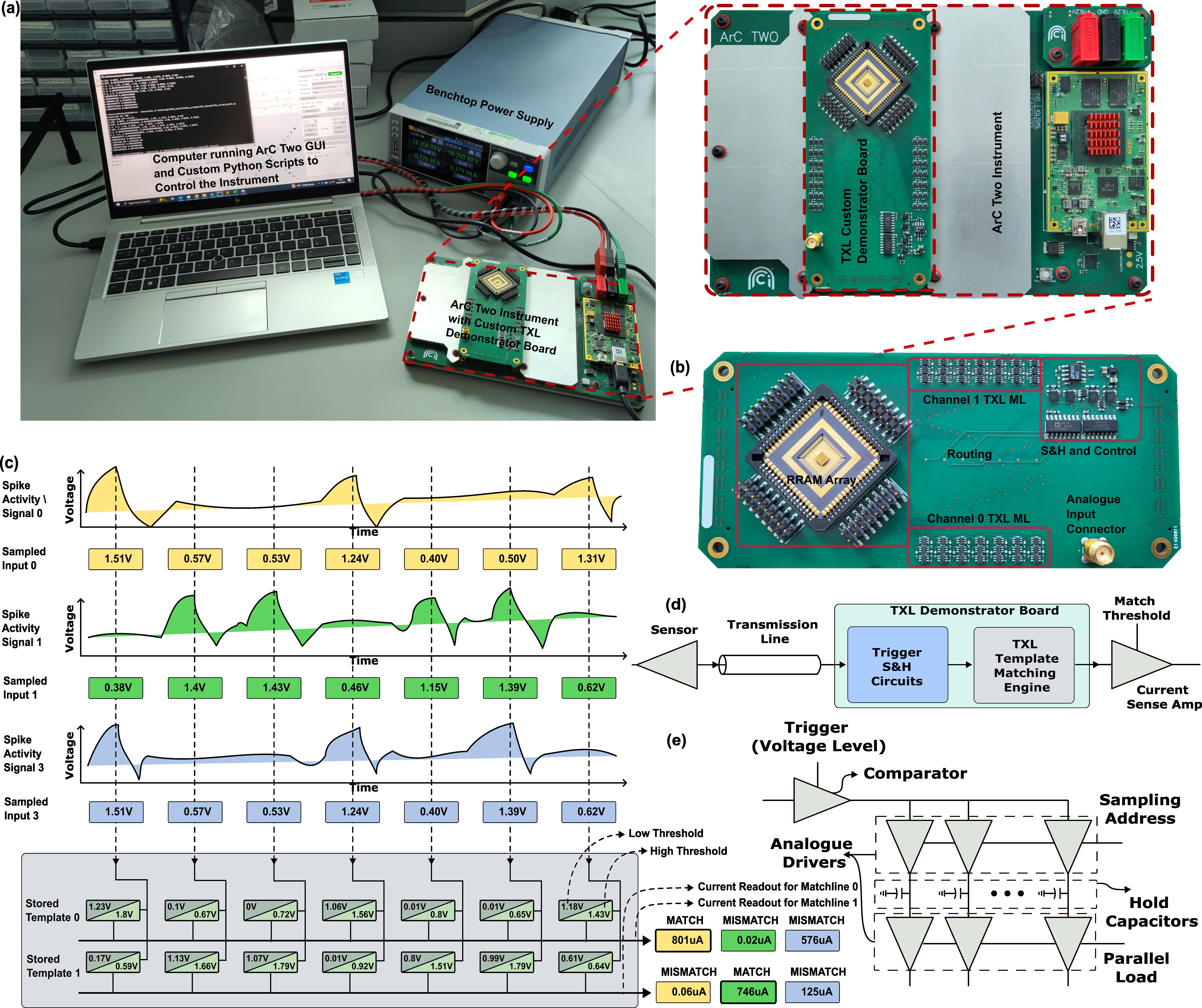}
        \caption{(a) The experimental setup of the TXL demonstrator board is shown. The TXL demonstrator consist of two main parts: the RRAM array and the CMOS components implementing the TXL circuits, the Sample \& Hold, Trigger system and control. The demonstrator is attached to the ArC TWO instrument. A benchtop power supply is used to provide the main voltage sources of the instrument board. The TXL demonstrator board is powered from the ArC Two main board. The controls for the main instrumentation board and the TXL demonstrator board are provided through PC. The input is provided by and the output is readout by the instrumentation board through appropriate setup of the flexible SMUs. (b) A closer looks at the custom demonstrator board used to test real RRAM devices with the TXL circuit. (c) An example operation of the TXL demonstration board is shown. Three different sets of sampled signals are used to test the cases where a hit in the first template, a hit in the second template and no hit is computed. The sets of query inputs are captured through the trigger and sample and hold circuit. The results of each comparison are represented as current readout and the use of a software threshold policy classifies the result per matchline as match or mismatch depending if above or below a predefined threshold value. The threshold value can be arbitrarily define and in this experiment the value of $V_{TH} = 700uA$ was selected. (d) A concept diagram showcasing the expected application. The TXL array implement a template matching engine capable of near-sensor classification. (e) A pre-processing unit is responsible for triggering the S\&H circuit and the samples are provided to the array.}
        \label{fig:TXL_Demo_Board}
\end{figure*}

 In order to demonstrate the functionality of the proposed TXL circuit alongside real RRAM devices designed and fabricated through in-house processes, we have designed and implemented a custom demonstrator board, shown in Fig. \ref{fig:TXL_Demo_Board}(a). The demonstrator board is compatible with the instrumentation infrastructure and can be used to test the functionality of multiple TXL circuits connected in matchlines and assess their behaviour. For the experiment, we are using MOSFET-based COTS circuits in addition to RRAM arrays we are fabricating in-house. Custom Python software script is controlling the different operation modalities of the demonstrator board. The custom demonstrator board used in the experiment is shown in Fig. \ref{fig:TXL_Demo_Board}(b). \par 
 %==========================================================
 %==========================================================
 In Fig. \ref{fig:TXL_Demo_Board}(d) we showcase a concept application of TXL array as part of a bio-signal classification engine, such as in the case of spike sorting template matching. In the light blue coloured area is the demonstrator board circuits to be tested in hardware using real RRAM devices for the TXL array, while the surrounding functions (sense, pre-processing, readout classification) are implemented in software. Part of the hardware implementation for the demonstrator board was the triggering and sample and hold circuit (shown in Fig. \ref{fig:TXL_Demo_Board}(e)) which can be operated when using an example waveform. The triggering circuit uses a voltage threshold trigger to identify when it needs to start sampling the input waveform. The sample and hold circuit is getting loaded with equally timed samples of the input waveform. The control of the sampling through timers and additional logic is also implemented on-board. \par
  %==========================================================
 As shown in Fig. \ref{fig:TXL_Demo_Board}(c), we are capturing distinct samples of a continuous signal, storing each one in a sample and hold cell. During the parallel search operation, the sampled voltages (forming an analogue query vector) are supplied as inputs to the two TXL rows with each row representing a stored template. Depending on each stored template a rate of match is calculated in analogue signal domain based on the match events accumulating for each template. Practically, the matching rate for each comparison of the query input with each template appears as current per matchline. The current is measured through ArC TWO Instrumentation Board SMUs that have been flexibly configured through the versatile ArC TWO readout channel mapping. We can define a threshold on each matchline which if surpassed will indicate a match event for the query. Otherwise, a mismatch occurs between the query and the template. The decision on the matching threshold can be decided arbitrarily depending on the specific hardware configuration of the implemented array as well as the application in question. It is possible through the matching threshold to perform an exact or approximate matching operation. \par
%==========================================================
 Through this demonstrator board we are able to initialise our investigation into the dynamics of the TXL circuits as part of larger ACAM arrays. The demonstrator board has been designed to accommodate extensive design for testing capabilities by enabling the access to all important nodes of each TXL cell for probing and thus enables the full control, characterisation and testing of the circuits and RRAM devices under test. Hence, a faster characterisation and calibration to ACAM operation cycle can be performed, accelerating the experimental process. The current demonstrator board enabled us to extract useful information to guide our approach when integrating a high number of RRAM devices as Back-End-Of-Line (BEOL) integration towards developing hybrid RRAM-CMOS ICs. \par

%===================================================================
\section{Circuit Simulations}\label{sec:sims}
\subsection{6T2R2M Cell Performance Analysis}
To assess integrability estimate the power consumed by the circuit, an integrated circuit model using the Cadence Virtuoso environment and a commercially available 180nm CMOS technology product development kit (PDK) was developed. This technology node includes planar MOSFETs, for either $\SI{1.8}{\volt}$ or $\SI{5}{\volt}$. For simulating the RRAM devices, we used a RRAM model based on the work of \cite{memModelPaper}. As integrated design provides much greater control of the specifics of the transistors used, three configurations of the proposed circuit were designed using different geometry parameters for employed the MOSFETs devices in the two hybrid RRAM-CMOS inverters (as shown in Table \ref{tab:FETsizes}): (1) with minimum size input transistors, (2) with wide input transistors, and (3) with minimum size native NMOS and conventional PMOS input transistors. In all cases the two transistors on the output branch were minimum size. All MOSFET components used for the showcased Spectre-SPICE simulations are $\SI{1.8}{\volt}$ devices. While it may be possible to replace the balancing resistors with static memristors, the assumption is that polysilicon resistors will be used. This constraint makes the $\SI{1}{\mega\ohm}$ balancing resistors of the PCB model wildly impractical, due to the area required. The simulations shown here were conducted with $\SI{10}{\kilo\ohm}$ balancing resistors and memristors of $\SI{2.5}{\kilo\ohm}$ to $\SI{250}{\kilo\ohm}$. Note that the memristor model used has the specified conductance at $\SI{0.5}{\volt}$ and has conductance closer to a $\SI{1}{\kilo\ohm}$ to $\SI{100}{\kilo\ohm}$ resistor at the inflection point of the inverters. In addition to this, an extra gate was added to both transistors in the output stage, to allow the output to be disabled when the input was being adjusted or the circuit shut down. While the move to a higher current operating point strips the circuit of a significant portion of its range, it allows for the recovery of much of the sharpness that is now expected to be lost when implemented with memristors. It should be noted that a reduced range is not necessarily a problem, as the circuits before the CAM cell in the signal chain can be designed to output in the operating range of the CAM cell, and may even have their own headroom requirements that make a wider window unnecessary.\par
%===================================================================
\begin{table}[ht]
    \caption{Chart of transistor parameters in the three circuit configurations that were simulated in Cadence Virtuoso using a commercial 180nm technology.}
    \centering
    \resizebox{0.8\linewidth}{!}{    
        \begin{tabular}{|c||c|c|c|}
        \hline
             & Minimum & Wide & Native \\
        \hhline{|=||=|=|=|}
        NMOS Width & $\SI{220}{\nano\meter}$ & $\SI{1}{\micro\meter}$ & $\SI{420}{\nano\meter}$\\
        \hline
        NMOS Length & $\SI{180}{\nano\meter}$ & $\SI{180}{\nano\meter}$ & $\SI{500}{\nano\meter}$\\
        \hline
        PMOS Width & $\SI{220}{\nano\meter}$ & $\SI{1}{\micro\meter}$ & $\SI{220}{\nano\meter}$\\
        \hline
        PMOS Length & $\SI{180}{\nano\meter}$ & $\SI{180}{\nano\meter}$ & $\SI{180}{\nano\meter}$\\
        \hline
        \end{tabular}
    }
    \label{tab:FETsizes}
\end{table}

\subsection{Performance Assessment}
%===================================================================
To assess the matching window maximum width, static DC simulations of the circuits were run. The circuits were supplied at $\SI{1.8}{\volt}$ and the inputs swept from $\SI{0}{\volt}$ to the supply voltage, measuring the current measured at the output. This was repeated for geometrically spaced values of $\SI{2.5}{\kilo\ohm}$ to $\SI{250}{\kilo\ohm}$ for the memristor. In all cases the output was connected to a $\SI{100}{\kilo\ohm}$ dummy load. The window widths were calculated using the same differential method mentioned in Section \ref{sec:mes}.\par
%===================================================================
    \begin{figure}[ht]
        \centering
        \begin{subfigure}[b]{0.49\linewidth}
            \centering
            \includegraphics[width=\linewidth]{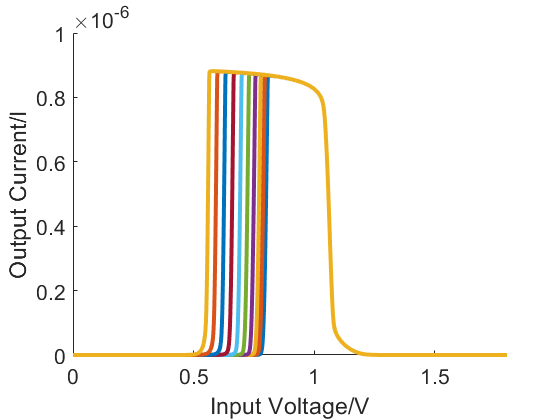}
            \caption{}
            \label{fig:memMinBot}
        \end{subfigure}
        \begin{subfigure}[b]{0.49\linewidth}
            \centering
            \includegraphics[width=\linewidth]{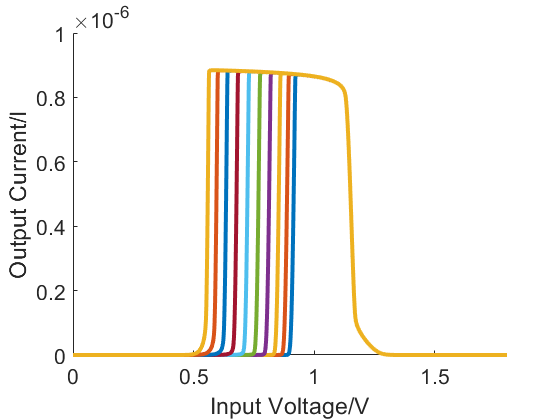}
            \caption{}
            \label{fig:memWideBot}
        \end{subfigure}
        \begin{subfigure}[b]{0.49\linewidth}
            \centering
            \includegraphics[width=\linewidth]{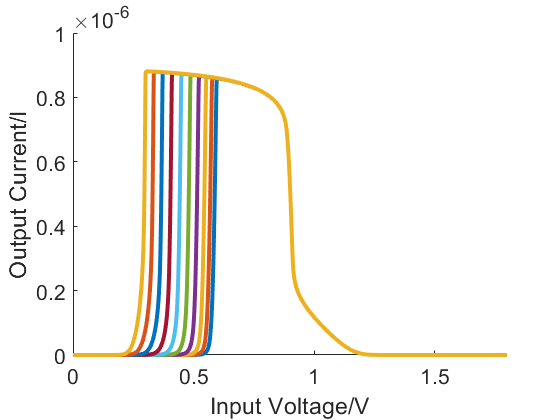}
            \caption{}
            \label{fig:memNatBot}
        \end{subfigure}
        \begin{subfigure}[b]{0.49\linewidth}
            \centering
            \includegraphics[width=\linewidth]{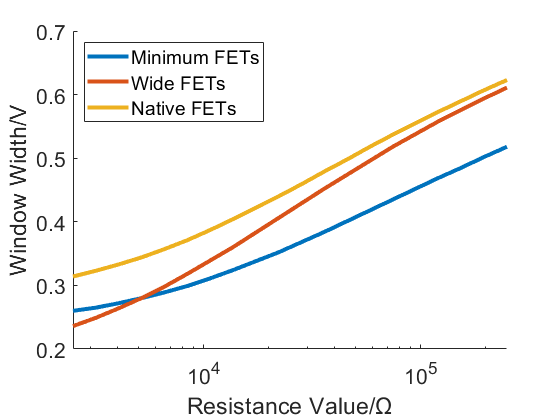}
            \caption{}
            \label{fig:memWidthBot}
        \end{subfigure}
        \caption{a) A graph of the IV characteristics of the minimum circuit, sweeping M1 from $\SI{2.5}{\kilo\ohm}$ to $\SI{250}{\kilo\ohm}$. b) An identical sweep of M1 with the wide circuit. c) An identical sweep of M1 with the native circuit. d) The window width of all three configurations as a function of M1.}
        \label{fig:memBot}
    \end{figure}
%===================================================================
In the circuit configuration with minimum size input transistors, the window width was drastically reduced. Furthermore, the range of values for the lower thresholds was roughly half the maximum window width. While the transitions were sharper, this reduced range represents a limitation in circuit's functionality, as a narrow window can only be set in a $\SI{100}{\milli\volt}$ range. The wide input stage circuit showcased better matching window configurability, losing less maximum width and retaining most of the threshold sweep. The sharpness of the transitions was also superior to the minimum width circuit, due to the higher gain of the input FETs. The native input stage circuit displayed a wider maximum window width than even the wide circuit, although the minimum size PMOS left it with a lower threshold range comparable to the minimum circuit. The matching window transitions in the native circuit were noticeably less sharp. As one might expect from the lower threshold voltages of native devices, the window of the native circuit was much closer to $\SI{0}{\volt}$ than the other configurations, which were almost centred in the supply range. In all circuits, plotting the window width as a function of the logarithm of M1 showed less non-linearity than lower threshold of the PCB model.\par
%===================================================================
    \begin{figure}[ht]
        \centering
        \begin{subfigure}[b]{0.49\linewidth}
            \centering
            \includegraphics[width=\linewidth]{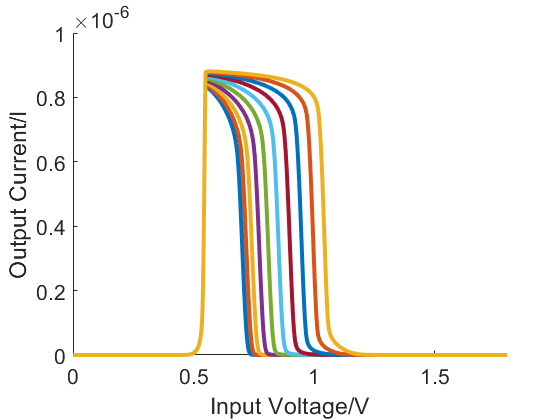}
            \caption{}
            \label{fig:memMinTop}
        \end{subfigure}
        \begin{subfigure}[b]{0.49\linewidth}
            \centering
            \includegraphics[width=\linewidth]{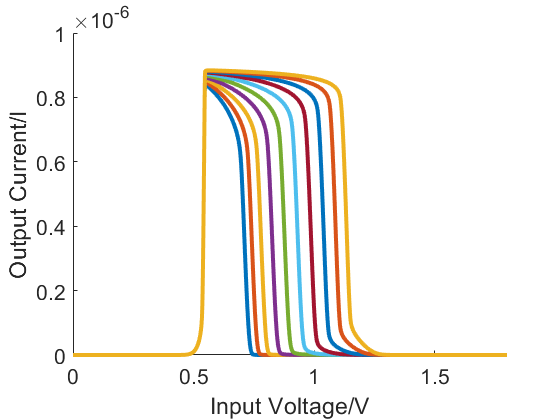}
            \caption{}
            \label{fig:memWideTop}
        \end{subfigure}
        \begin{subfigure}[b]{0.49\linewidth}
            \centering
            \includegraphics[width=\linewidth]{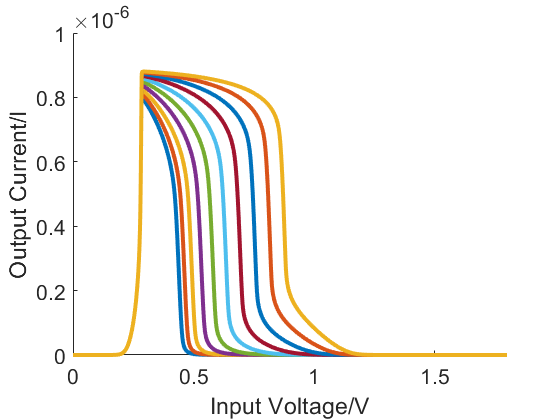}
            \caption{}
            \label{fig:memNatTop}
        \end{subfigure}
        \begin{subfigure}[b]{0.49\linewidth}
            \centering
            \includegraphics[width=\linewidth]{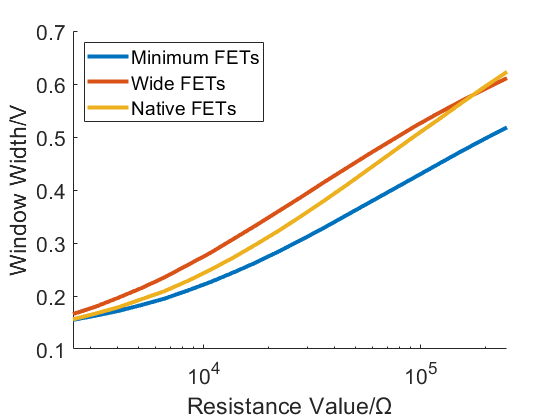}
            \caption{}
            \label{fig:memWidthTop}
        \end{subfigure}
        \caption{a) A graph of the IV characteristics of the minimum circuit, sweeping M2 from $\SI{2.5}{\kilo\ohm}$ to $\SI{250}{\kilo\ohm}$. b) An identical sweep of M2 with the wide circuit. c) An identical sweep of M2 with the native circuit. d) The window width of all three configurations as a function of M2.}
        \label{fig:memTop}
    \end{figure}
%===================================================================
The differences in the behaviour of the top threshold between the three circuit configurations were less pronounced. All three were able to bring the top threshold to within $\SI{100}{\milli\volt}$ of the bottom threshold. The only significant differences were the inability of the minimum width circuit to reach the maximum window widths of the other two, and the less defined transitions of the native circuit.\par
%===================================================================
\subsection{Energy Dissipation}

    \begin{figure}[ht]
        \centering
        \includegraphics[width=0.6\linewidth]{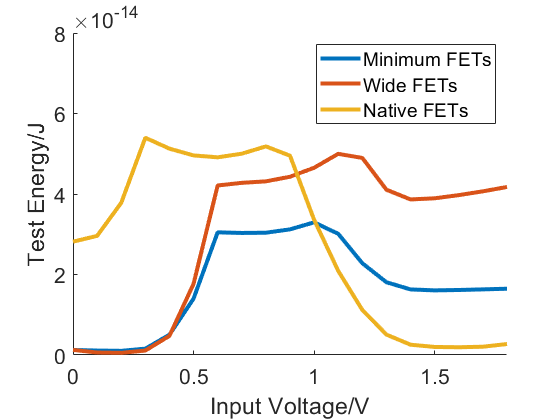}
        \caption{A graph of the test energy of a single simulated 6T2M2R ACAM cell with maximum window width, as a function of input voltage.}
        \label{fig:TXLenergy}
    \end{figure}
%===================================================================
To estimate the energy required to test a single sample, a transient analysis simulation was run. In this test, the circuit began with the supply energised, the output disabled through the gates added to the output FETs, and the input parked at either supply or ground. The minimum and wide circuits were parked at ground and the native circuit parked at the supply voltage. The input was brought to the test value and held there for $\SI{2.35}{\nano\second}$ to settle. The output enable signal was then pulsed for $\SI{450}{\pico\second}$. $\SI{200}{\pico\second}$ after the enable pulse ends, the input is returned to its parked state. The energy was calculated by integrating the sum of the instantaneous power at the input, the enable input, and the supply. As might be expected for an analysis including the output current, the test energy of a hit was the peak energy in all cases. It is worth noting that the miss energy on the opposite side of the window to the parked state was substantially higher than the miss energy on the same side of the window. This observation is probably because bringing the input signal to the test voltage and back causes the input to pass through both input inverter's crossover points. This would cause a small surge of current each time, resulting in the observed plateau.\par
%===================================================================
\begin{table}[ht]
\caption{Energies calculated at typical input stimulus voltages for several process corners in both hit and miss cases. Note that the difference in hit/miss voltage between the circuit variants make this data less useful for comparing the variants. It is more intended to show the response of each circuit to process variation and temperature.}
    \centering
        \centering
        \resizebox{0.8\linewidth}{!}{    
            \begin{tabular}{|c||c|c|c|}
            \hline
                 & Minimum & Wide & Native \\
            \hhline{|=||=|=|=|}
            Hit Voltage & $\SI{900}{\milli\volt}$ & $\SI{900}{\milli\volt}$ & $\SI{600}{\milli\volt}$\\
            \hline
            $\SI{25}{\celsius}$ & $\SI{31.21}{\femto\joule}$ & $\SI{44.29}{\femto\joule}$ & $\SI{49.11}{\femto\joule}$\\
            \hline
            $\SI{37}{\celsius}$ & $\SI{31.70}{\femto\joule}$ & $\SI{44.97}{\femto\joule}$ & $\SI{49.75}{\femto\joule}$\\
            \hline
            Fast/Fast Corner & $\SI{40.09}{\femto\joule}$ & $\SI{51.18}{\femto\joule}$ & $\SI{62.80}{\femto\joule}$\\
            \hline
            Slow/Slow Corner & $\SI{23.90}{\femto\joule}$ & $\SI{37.74}{\femto\joule}$ & $\SI{38.26}{\femto\joule}$\\
            \hhline{|=||=|=|=|}
            Miss Voltage & $\SI{1.3}{\volt}$ & $\SI{1.5}{\volt}$ & $\SI{1.2}{\volt}$\\
            \hline
            $\SI{25}{\celsius}$ & $\SI{18.07}{\femto\joule}$ & $\SI{38.93}{\femto\joule}$ & $\SI{11.19}{\femto\joule}$\\
            \hline
            $\SI{37}{\celsius}$ & $\SI{18.61}{\femto\joule}$ & $\SI{39.26}{\femto\joule}$ & $\SI{12.36}{\femto\joule}$\\
            \hline
            Fast/Fast Corner & $\SI{24.54}{\femto\joule}$ & $\SI{40.61}{\femto\joule}$ & $\SI{24.46}{\femto\joule}$\\
            \hline
            Slow/Slow Corner & $\SI{15.91}{\femto\joule}$ & $\SI{37.43}{\femto\joule}$ & $\SI{4.325}{\femto\joule}$\\
            \hline
            \end{tabular}
        }
    \label{tab:energyTable}
\end{table}
%===================================================================
\begin{table}[ht]
    \caption{Maximum matching window width of the proposed ACAM cell at several 180nm CMOS technology process corners.}
    \centering
        \resizebox{\linewidth}{!}{    
            \begin{tabular}{|c||c|c|c|}
            \hline
            Maximum Window Width at & Minimum & Wide & Native \\
            \hhline{|=||=|=|=|}
            $\SI{25}{\celsius}$ & $\SI{517}{\milli\volt}$ & $\SI{609}{\milli\volt}$ & $\SI{622}{\milli\volt}$\\
            \hline
            $\SI{37}{\celsius}$ & $\SI{523}{\milli\volt}$ & $\SI{618}{\milli\volt}$ & $\SI{627}{\milli\volt}$\\
            \hline
            Fast/Fast Corner & $\SI{634}{\milli\volt}$ & $\SI{706}{\milli\volt}$ & $\SI{727}{\milli\volt}$\\
            \hline
            Slow/Slow Corner & $\SI{399}{\milli\volt}$ & $\SI{513}{\milli\volt}$ & $\SI{517}{\milli\volt}$\\
            \hline
            \end{tabular}
        }
    \label{tab:hitMissTable}
\end{table}
%===================================================================
Further simulation data was gathered for relevant process corners. A corner is a set of environmental or process conditions that might affect circuit performance. Each of the circuits was simulated at four corner conditions: room temperature, body temperature, PMOS/NMOS above specified gain (Fast/Fast), and PMOS/NMOS below specified gain (Slow/Slow). A hit and a miss voltage were chosen for each circuit, to give estimates for a typical use case. The variation with temperature was minimal, but the effect of process variation was substantial, with a Fast/Fast hit dissipating almost twice the energy of a Slow/Slow hit.\par
%===================================================================
\subsection{Process Variation Effects}
\begin{figure}[ht]
        \centering
        \begin{subfigure}[b]{\linewidth}
        \centering
            \includegraphics[width=\linewidth]{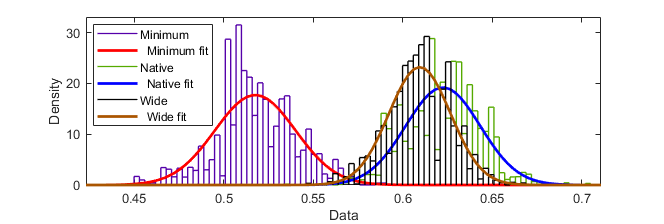}
            \caption{}
            \label{fig:MChiat}
        \end{subfigure}
        \begin{subfigure}[b]{0.74\linewidth}
            \centering
            \resizebox{0.7\linewidth}{!}{    
                \begin{tabular}{|c||c|c|c|}
                \hline
                 & Minimum & Wide & Native \\
                \hhline{|=||=|=|=|}
                $\mu$ & $\SI{518}{\milli\volt}$ & $\SI{610}{\milli\volt}$ & $\SI{623}{\milli\volt}$\\
                \hline
                $\sigma$ & $\SI{22.5}{\milli\volt}$ & $\SI{17.2}{\milli\volt}$ & $\SI{20.8}{\milli\volt}$\\
                \hline
                \end{tabular}
            }
            \caption{}
            \label{tab:fitTable}
        \end{subfigure}
        \caption{a) A histogram plot and fit of a 250 point MC simulation investigating the maximum window width when process variations are considered. The process variations affect only the MOSFET and polysilicon balancing resistor components of the circuit design. b) A table of the fit parameters $\mu$ and $\sigma$, assuming a normal distribution.}
        \label{fig:MCfig}
    \end{figure}

To obtain a clearer picture of the impact of process variation, a Monte Carlo (MC) analysis was conducted in Cadence Virtuoso simulation environment. This approach uses a large number of simulations with randomised process variation to provide a statistical model of the expected impact. In this test, the characteristic being monitored is the maximum window width, calculated using the same differential method used earlier. To reinforce the statistical importance of our results, 250 point simulations were run for each circuit configuration. In Fig. \ref{fig:MCfig} the aggregate results for the MC simulation are exhibited. \par
%===================================================================
%===================================================================
The resulting histograms of window width show a significant level of variation in the minimum and native circuits. If an error in fabrication causes a gate element to be $\SI{20}{\nano\metre}$ narrower than intended, the proportional error will be much larger on a minimum size FET than a wider one, so greater variance on the minimum size circuit is to be expected. Why the native circuit has slightly greater variance is not entirely clear, as the larger minimum size of the native FETs should result in marginally less variance, but the low doping level of the native devices likely makes them more vulnerable to fabrication errors there. The wide circuit showed the least variance, both in absolute terms and in proportion of average maximum window width.\par
%===================================================================

\section{Conclusions \& Discussion}\label{sec:disc}
In this work, we are proposing a circuit for template matching operation based on RRAM technology. RRAM devices are employed to map analogue weights into configurable thresholds for ACAM applications. The circuit considered in this work demonstrates a highly controllable and flexible analogue content addressable memory cell with acceptable power dissipation. As part of the energy-efficient design methodology we followed for this work, the circuit uses a current-mode charging scheme for the matchline instead of a more conventional precharged matchline with each cell discharging if there is a mismatch. Additionally, this enables the generation of an analogue output per template, which can be used to calculate the distance between the query input and the stored template. The proposed circuit can be used in many different neuro-inspired circuit and systems, such as a wide variety of ANNs, but the main focus of this circuit to perform template matching operations in a highly configurable and noise-tolerant approach makes it a promising candidate for in-sensor or near-sensor analogue signal processing at the edge, such as for the case of near-sensor neural spike sorting systems. Compared to the reconfigurable logic gates this work was developed from \cite{TXLpaper}, this work demonstrates a greater degree of controllability for a similar device count and area. The match window can retain its configurability and sharp selection even with a smaller conductance ratios compared to prior work thus enabling a easier integration with a wide variety of potential RRAM technologies. Experimental testing with different RRAM technologies to assess the performance of the circuit under different operational conditions is planned as future work. This is possible due to the design of technology-agnostic PCB test and demonstrator board as shown in this work. \par
%===================================================================
In future work, we are planning to further improve the current design and work towards a prototype integrated solutions to test a larger number of RRAM devices and TXL circuits using the existing framework. Additionally, it is worth noting that as the ratio between dynamic and balancing elements is the only control parameter that has a significant effect, $R_{M1}$ and $R_{1}$ could swap positions. This would allow for both thresholds to have identical control methods, at the trivial cost of inverting the control of one threshold. With both dynamic elements connected to ground, the devices could be written through the diode structures formed by the triple well of the neighbouring NMOS, although this method has yet to be demonstrated. The power and area requirements for the circuit could be further improved by switching to a smaller technology node, such as $\SI{22}{\nano\metre}$. Such circuit miniaturisation regarding its CMOS part will require a careful investigation in order to ensure the continuity of co-integration viability with RRAM technologies, such as the $TiO_{x}$-based RRAM devices used in this work. Through the demonstrator board we showed how these low power window comparators might be arrayed to perform near sensor classification tasks. These experiments enabled us to further mature our design efforts for a higher level of integration of RRAM with CMOS through BEOL monolithic integration. \par
%===================================================================

%%%%%%%%%%%%%%%%%%%%%%%%%%%%%%%%%%%%%%%%%%%%%%%%%%%%%%%%%%%%%%%%%%%%%%%%%%%%%%%%

%%%%%%%%%%%%%%%%%%%%%%%%%%%%%%%%%%%%%%%%%%%%%%%%%%%%%%%%%%%%%%%%%%%%%%%%%%%%%%%%

%%%%%%%%%%%%%%%%%%%%%%%%%%%%%%%%%%%%%%%%%%%%%%%%%%%%%%%%%%%%%%%%%%%%%%%%%%%%%%%%

%%%%%%%%%%%%%%%%%%%%%%%%%%%%%%%%%%%%%%%%%%%%%%%%%%%%%%%%%%%%%%%%%%%%%%%%%%%%%%%%

\bibliography{ALL_library_3}

\end{document}